\documentclass[times,doublespace]{cpeak2}

\usepackage{moreverb}

\usepackage[dvips,colorlinks,bookmarksopen,bookmarksnumbered,citecolor=red,urlcolor=red]{hyperref}

\newcommand\BibTeX{{\rmfamily B\kern-.05em \textsc{i\kern-.025em b}\kern-.08em
T\kern-.1667em\lower.7ex\hbox{E}\kern-.125emX}}

\usepackage{amsmath}

\newcommand{\refbr}[1]{(\ref{#1})}
\newcommand{\inI}[2]{= #1, \ldots, #2}
\newcommand{\beq}{\begin{equation}}
\newcommand{\eeq}{\end{equation}}

\begin{document}

\runningheads{A.Karbowski}{Remarks to  "Control system for reducing energy consumption...}

\title{
Remarks to the
paper  "Control system for reducing energy consumption in backbone
computer network" from "Concurrency and Computation: Practice and Experience" journal
}

\author{Andrzej Karbowski\corrauth}

\address{
Institute of Control and Computation Engineering, Warsaw University of Technology, ul. Nowowiejska 15/19,
00-665 Warsaw, Poland
}

\corraddr{E-mail:A.Karbowski@elka.pw.edu.pl}

\begin{abstract}
This paper indicates two errors in the formulation of the main
optimization model in the article "Control system for reducing energy consumption in backbone
computer network" by Niewiadomska-Szynkiewicz et al. and shows how
to fix them.

\end{abstract}


\maketitle

\vspace{-6pt}

\section{Introduction, model from the source paper}
\vspace{-2pt}

The paper \cite{ens}  considers a backbone computer network formed by
 routers labeled with  $r=1,\ldots,R$,  line cards labeled with $c=1,\ldots,C$,  communication ports labeled with $p=1,\ldots,P$ and directed links labeled with $e=1,\ldots,E$. The hierarchical representation of a router is assumed, i.e., each router is equipped with a number of line cards and each card contains a number of communication ports. All network components can operate in different energy states.  Two ports connected by the same link $e$ are in the same state $k=1,\ldots,K$, which at the expense
 of the power consumption $\xi_{ek}$ allows for the transmission throughput  $M_{ek}$.
  $W_c$ and $T_r$ are fixed power cost components associated to utilizing, respectively, a card $c$ and a router $r$.

 All demands imposed on the network are labeled with $d=1,\ldots,D$ and characterized  by the source node $s_d$, the destination node $t_{d}$ and a given volume $V_d$, ensuring end-to-end Quality of Service (QoS).

The topology of the network is described by four matrices of binary indicators: $l_{cp}, g_{rc}, a_{ep}, b_{ep}$, whether, respectively: port $p$ belongs to the card $c$, card $c$ belongs to the router $r$,
link $e$ is outgoing from the port $p$ and link $e$ is incoming to the port $p$.

The decision variables are two vectors of binary indicators $x_{c}, z_{r}$ -
whether the card $c$ or the router $r$ is used for data transmission and two incidence matrices
with elements: $y_{ek}$ - whether the link $e$ is in the state $k$ and $u_{ed}$ - whether the
demand $d$ uses the link $e$.

%
The energy saving optimization problem which aim is to minimize the total power utilized by network components while ensuring the QoS requirements is formulated in \cite{ens} as follows:

\begin{equation}
\min_{x_c,y_{ek},z_r,u_{ed}} \left[ F_{LN} =
\sum_{e=1}^{E}\sum_{k=1}^{K}\xi_{ek}y_{ek} + \sum_{c=1}^{C}W_{c}x_{c}
 + \sum_{r=1}^R T_{r}z_{r}
\right] ,
\label{c1a}
\end{equation}
subject to the constraints:

\beq
\forall_{\substack{d=1,\ldots,D,\\
	c=1,\ldots,C}}  \quad \sum_{p=1}^{P}l_{cp}\sum_{e=1}^{E}a_{ep}u_{ed} \leq x_{c},
\label{c2a}
\eeq
\beq
\forall_{\substack{d=1,\ldots,D,\\
	c=1,\ldots,C}}  \quad \sum_{p=1}^{P}l_{cp}\sum_{e=1}^{E}b_{ep}u_{ed} \leq x_{c},
\label{c3a}
\eeq
\beq
\forall_{\substack{r=1,\ldots,R,\\
	c=1,\ldots,C}}  \quad g_{rc}x_{c} \leq z_{r},
\label{c4a}
\eeq
\beq
\forall_{\substack{e=1,\ldots,E}}  \quad \sum_{k=1}^{K}y_{ek} \leq 1,
\label{c5a}
\eeq
\beq
\forall_{\substack{d=1,\ldots,D,\\
	p=s_{d}}} \quad
            \sum_{e=1}^{E}a_{ep}u_{ed}
         -  \sum_{e=1}^{E}b_{ep}u_{ed}
         = 1,
\label{c6a}
\eeq
\beq
\forall_{\substack{d=1,\ldots,D,\\
    p\neq t_d, p\neq s_d}} \quad
            \sum_{e=1}^{E}a_{ep}u_{ed}
         -  \sum_{e=1}^{E}b_{ep}u_{ed}
         = 0,
\label{c7a}
\eeq
\beq
\forall_{\substack{d=1,\ldots,D,\\
	p = t_{d}}} \quad
             \sum_{e=1}^{E}a_{ep}u_{ed}
         - \sum_{e=1}^{E}b_{ep}u_{ed}
         = -1,
\label{c8a}
\eeq
\beq
\forall_{\substack{e=1,\ldots,E}}  \quad \sum_{d=1}^{D}V_{d}u_{ed} \leq \sum_{k=1}^{K}M_{ek}y_{ek}.
\label{c9a}
\eeq
In the above formulation the constraints \refbr{c2a}-\refbr{c4a} determine the number of routers and cards used for data transmission, the conditions \refbr{c5a} assure that each link can be in one energy-aware state. The constrains \refbr{c6a}-\refbr{c8a} are formulated  according to the 1st Kirchhoff's law applied for source, transit
and destination routers, and the constraint \refbr{c9a} assures, that the flow will not exceed the capacity $M_{ek}$ of a given link.

Unfortunately, when the author tried to use the formulation \refbr{c1a}-\refbr{c9a} to solve
a test problem, he got wrong results. Having spent quite a lot of time he found the reason - errors in the model.

\section{Errors in the model and their correction}

There are two errors in the formulation \refbr{c1a}-\refbr{c9a}:
\begin{enumerate}
\item Flow conservation equations \refbr{c6a}-\refbr{c8a} are incorrectly written.
    The reason is, that in a backbone computer network routers are nodes, not ports, as it is done
    in Eqs. \refbr{c6a}-\refbr{c8a}, where $p$ is a fixed parameter.
    A port is only a labeled  input to a router (node), where the switch of routes is done.
    Moreover, every port in a router
    can be an input or an output for signals and the summation over them and, at the same time,
    over all links outcoming and incoming to the router, should be performed.\\

   \item Despite the announcement at the beginning of the section 5. of the article \cite{ens}:  "We
assume that at a given time, instant two ports connected by the $e$-th link are in the same state k.",
there are no equations ensuring it. The conditions \refbr{c9a} expressing the energy used by links are
formulated independently for all links. \\
The above assumption is natural in computers networks and it should be reflected in a good  model.
\end{enumerate}

To fix the two errors mentioned above it is proposed:
\begin{description}
\item[Ad.1.] To replace  the three equations \refbr{c6a}-\refbr{c8a}  by
        the following one:
\beq
\forall_{\substack{d=1,\ldots,D,\\
        r=1,\ldots,R}}
            \sum_{c=1}^C g_{rc} \sum_{p=1}^P l_{cp} \sum_{e=1}^{E} \left(a_{ep}
         -   b_{ep} \right)u_{ed}
         =
\left\{\begin{array}{ll}
          1 & \,\; r=s_{d},\\
         -1 & \,\; r = t_{d},\\
          0 & \, \mbox{otherwise}.
         \end{array}\right.
\eeq
In this equation the summation is done across every router $r=1,\ldots,R$ for every demand $d=1,\ldots,D$.
All links connected to the router $r$ are taken into account owing to the
summations:
\beq
\sum_{c=1}^C \sum_{p=1}^P \sum_{e=1}^E g_{rc} l_{cp} a_{ep} u_{ed}
\eeq
for the outgoing traffic
and
\beq
\sum_{c=1}^C \sum_{p=1}^P \sum_{e=1}^E g_{rc} l_{cp} b_{ep} u_{ed}
\eeq
for the incoming traffic.
\item[Ad.2.] To augment the conditions \refbr{c9a} with equality constraints assuring
    that the energy level in both links of every edge is the same.
    They are as follows:
    \beq
    \forall_{\substack{p = 1,\ldots,P\\
             k=1,\ldots,K}} \sum_{e=1}^E a_{ep} y_{ek} = \sum_{e=1}^E b_{ep} y_{ek}
             \label{dodat.rown}
    \eeq
    Since in equation \refbr{dodat.rown} indices $p$ and $k$ are fixed, with the assumptions
    taken, for a given port $\bar{p}$ there is only one combination of links $e_1, e_2 \in 1,\ldots,E$, such that:
    \beq
    a_{e_1\bar{p}}=b_{e_2\bar{p}}=1, \forall_{e\neq e_1} a_{e,\bar{p}}=0, \forall_{e \neq e_2} b_{e,\bar{p}}=0.
    \eeq
    Taking this into account from equation \refbr{dodat.rown} we will get for all $k=1,\ldots,K$:
    \beq
    y_{e_1k}=y_{e_2k}
    \eeq
    The same reasoning may be repeated for the opposite port $\bar{\bar{p}}$ of the edge, such that:
    \beq
    b_{e_1\bar{\bar{p}}}=a_{e_2\bar{\bar{p}}}=1, \forall_{e\neq e_1} b_{e,\bar{\bar{p}}}=0, \forall_{e \neq e_2} a_{e,\bar{\bar{p}}}=0.
    \eeq
    It  means, that  the energy level will be the same in the edge formed of links $e_1$ and $e_2$.
\end{description}

\section{The final formulation of the problem}

Summing up, the final formulation of the energy saving backbone
 network control  problem will be as follows \footnote{Under minimization operator
all sets of indices of the arguments of optimization have been added. They were incorrectly omitted in \cite{ens}
(see  Eq. \refbr{c1a}).}

\begin{equation}
\min_{\substack{x_c,y_{ek},z_r,u_{ed}\\
c \inI{1}{C}; e \inI{1}{E}; k \inI{1}{K}\\
r \inI{1}{R}; d \inI{1}{D}}} \left[ F_{LN} =
\sum_{e=1}^{E}\sum_{k=1}^{K}\xi_{ek}y_{ek}
+ \sum_{c=1}^{C}W_{c}x_{c} +
\sum_{r=1}^RT_{r}z_{r}  \right] ,
\label{c1N}
\end{equation}
subject to the constraints:
\beq
\forall_{\substack{d=1,\ldots,D,\\
	c=1,\ldots,C}}  \quad \sum_{p=1}^{P}l_{cp}\sum_{e=1}^{E}a_{ep}u_{ed} \leq x_{c},
\label{c2N}
\eeq
\beq
\forall_{\substack{d=1,\ldots,D,\\
	c=1,\ldots,C}}  \quad \sum_{p=1}^{P}l_{cp}\sum_{e=1}^{E}b_{ep}u_{ed} \leq x_{c},
\label{c3N}
\eeq
\beq
\forall_{\substack{r=1,\ldots,R,\\
	c=1,\ldots,C}}  \quad g_{rc}x_{c} \leq z_{r},
\label{c4N}
\eeq
\beq
\forall_{\substack{e=1,\ldots,E}}  \quad \sum_{k=1}^{K}y_{ek} \leq 1,
\label{c5N}
\eeq
\beq
\forall_{\substack{d=1,\ldots,D,\\
        r=1,\ldots,R}}
            \sum_{c=1}^C g_{rc} \sum_{p=1}^P l_{cp} \sum_{e=1}^{E} \left(a_{ep}
         -   b_{ep} \right)u_{ed}
         =
\left\{\begin{array}{ll}
          1 & \,\; r=s_{d},\\
         -1 & \,\; r = t_{d},\\
          0 & \, \mbox{otherwise},
         \end{array}\right.
\label{c6-8N}
\eeq
\beq
\forall_{\substack{e=1,\ldots,E}}  \quad \sum_{d=1}^{D}V_{d}u_{ed} \leq \sum_{k=1}^{K}M_{ek}y_{ek},
\label{c9N}
\eeq
\beq
    \forall_{\substack{p = 1,\ldots,P\\
             k=1,\ldots,K}} \sum_{e=1}^E a_{ep} y_{ek} = \sum_{e=1}^E b_{ep} y_{ek},
    \eeq

 \beq
x_c, z_r \in \{0,1\} \;\, c \inI{1}{C}; r \inI{1}{R};
\label{bina12}
\eeq
\beq
y_{ek},u_{ed} \in \{0,1\}\;\, e \inI{1}{E}; k \inI{1}{K}; d \inI{1}{D}.
\label{bina22}
\eeq

\section{Conclusions}

The indicated errors in the formulation of the main optimization model of
energy saving backbone network control presented in \cite{ens} concern flow balance equations for all nodes
and the requirement of the same energy level in two directional links
connecting the same ports. To make this model correct it was necessary to modify the flow balance equations, treating routers
as nodes (instead of ports as it is in \cite{ens}) and to add equality constrains
on the levels of power consumption in two links incoming to and outgoing from the same port.
Without these changes the model does not describe well the dependencies between different
components of the backbone computer network, including routers, cards, ports and links and is not fully useful.

\section*{Acknowledgments}
Before the publication in Arxiv.org this article was submitted to "Concurrency and Computation: Practice and Experience" journal on 10 Aug. 2016. However, as a consequence, instead of it an erratum \cite{ens2} to the paper \cite{ens} by its authors,  changing the model into the one presented in \cite{niewFGCS}, was published
on 25 Dec. 2016.
 The author wants to thank for mentioning him in the Acknowledgment section of \cite{ens2}, but in his
 opinion the newer model, stemming from \cite{niewFGCS}, has still errors.
 They have been described and fixed in \cite{jaFGCS}, before the first response from "Concurrency and Computation: Practice and Experience".


\begin{thebibliography}{1}
\bibitem{ens}Niewiadomska-Szynkiewicz E, Sikora A, Arabas P, Ko{\l}odziej J. Control system for
reducing energy consumption in backbone computer network. Concurrency Computat.: Pract. Exper. 2013; 25:1738–-1754
\bibitem{ens2}Niewiadomska-Szynkiewicz E, Sikora A, Arabas P, Ko{\l}odziej J. ERRATUM
Control system for reducing energy consumption in backbone
computer network. Concurrency Computat.: Pract. Exper. 2016; 28:4557
\bibitem{niewFGCS}Niewiadomska-Szynkiewicz E, Sikora A, Arabas P, Kamola M, Mincer M, Ko{\l}odziej J.  Dynamic power management in energy-aware computer networks and data intensive computing systems;
  Future Generation Computer Systems 2014,  37:284--296
\bibitem{jaFGCS} Karbowski A. Correction to the article "Dynamic power management in energy-aware computer networks and data intensive computing systems" published in "Future Generation Computer Systems" journal.
    	arXiv:1610.02551 [cs.DC] \url{https://arxiv.org/abs/1610.02551} 2016. 
\end{thebibliography}
\end{document}